\documentclass[a4paper,12pt,twoside]{cpc-hepnp}
\usepackage{multicol}
\usepackage{graphicx}
\usepackage{booktabs}
\usepackage{amssymb,bm,mathrsfs,bbm,amscd}
\usepackage[tbtags]{amsmath}
\usepackage{lastpage}
\usepackage{color}
\usepackage{enumerate}
\usepackage{lineno}
\usepackage{url}
\usepackage[english]{babel}

\begin{document}
%\begin{CJK*}{GBK}{song}

\fancyhead[c]{\small Submitted to Chinese Physics C} \fancyfoot[C]{\small \thepage}

\title{Onsite data processing and monitoring for the Daya Bay Experiment}

\author{
      LIU Ying-Biao$^{1;1)}$\email{liuyb@ihep.ac.cn}%
\quad HE Miao$^{1;2)}$\email{hem@ihep.ac.cn}%
\quad LIU Bei-Jiang$^{1,2,3}$%
\quad WANG Meng$^{4}$\\%
\quad MA Qiu-Mei$^{1}$%
\quad QI Fa-Zhi$^{1}$
\quad ZENG Shan$^{1}$%
}
\maketitle

\address{
$^{1}$ Institute of High Energy Physics, Chinese Academy of Sciences, Beijing 100049, China\\
$^{2}$ Chinese University of Hong Kong, Hong Kong\\
$^{3}$ Department of Physics, The University of Hong Kong, Pokfulam, Hong Kong\\
$^{4}$ Shandong University, Jinan 250100, China\\
}

\begin{abstract}
The Daya Bay Reactor Neutrino Experiment started running on September 23, 2011. The offline computing environment, consisting of 11 servers at Daya Bay, was built to process onsite data. With current computing ability, onsite data processing is running smoothly. The Performance Quality Monitoring system (PQM) has been developed to monitor the detector performance and data quality. Its main feature is the ability to efficiently process multi-data-stream from three experimental halls. The PQM processes raw data files from the Daya Bay data acquisition system, generates and publishes histograms via a graphical web interface by executing the user-defined algorithm modules, and saves the histograms for permanent storage. The fact that the whole process takes only around 40 minutes makes it valuable for the shift crew to monitor the running status of all the sub-detectors and the data quality.
\end{abstract}
\begin{keyword}
multi-data-stream, data quality, Daya Bay
\end{keyword}

\begin{pacs}
29.40.Mc, 28.50.Hw, 13.15.+g
\end{pacs}

\footnotetext[0]{\hspace*{-2em}\small\centerline{\thepage\ --- \pageref{LastPage}}}

\begin{multicols}{2}

\section{Introduction}
\par
The Daya Bay experiment is designed to determine precisely the neutrino mixing angle $\theta_{13}$ with a sensitivity better than 0.01 in the parameter $\sin^{2}\rm 2\theta_{13}$ at 90\% confidence level. Three experimental halls (EHs) were built to contain eight functionally identical antineutrino detectors (ADs)~\cite{ad12}. Before October 18, 2012, two ADs were located in EH1 and one in EH2 (the near halls). Three ADs were positioned in the far hall, EH3. After the installation of the last two ADs, 2 and 4 ADs are located in each near hall and the far hall, respectively. In each experimental hall there are also muon detectors, including the inner (IWS), outer (OWS) water shields, and a resistive plate chamber (RPC) tracker. With the 6-AD configuration, the Daya Bay collaboration reported the first observation of non-zero $\sin^22\theta_{13}$~\cite{prl}, and has recently updated the result of $\sin^22\theta_{13}=0.089\pm 0.010({\rm stat.})\pm0.005({\rm syst.})$~\cite{cpc} in a three-neutrino framework.

During physics data taking, events from different experimental halls are separately assembled in raw data format and stored into persistent files on the online disk array by the data acquisition system (DAQ). Event rates of 8-AD configuration for EH1, EH2, EH3 are about 1.2 kHz, 1.0 kHz, and 600 Hz, respectively. The total number of raw data files for each day is about 320, approximately 1 GB volume per file. The online histogram presenter of the DAQ provides the hit-maps for sub-detectors. In addition, high-level histograms are necessary to be produced for monitoring the running status of all sub-detectors and the data quality (DQ).

The Performance Quality Monitoring system (PQM) is developed to utilize the offline software to process multi-data-stream from three experimental halls, generates more user-defined histograms, including some high-level ones containing the calibration and reconstruction information, and thus plays an important role in offline monitoring for the experiment. Even with the total event rate of around 3 kHz, the PQM can handle the whole process around 40 minutes.

In this article, we present the offline computing environment at Daya Bay in Section 2. In Section 3, the PQM design is outlined. The performance and the information display of the PQM are shown in Section 4. We give a short summary in the last section.

\section{Offline computing environment at Daya Bay}

\subsection{Offline computers}

There are 11 servers in the offline computing \mbox{environment} at Daya Bay, including a file server, a data transfer server, an offline database server, a web server, two PQM servers (pqm1-2), and the rest of 5 servers forming user farms (farm1-5), as shown in Fig.~\ref{fig:oe}. The pqm1 is usually working as the PQM server, in which the main PQM control script is running, and the pqm2 is a backup machine. The PQM servers together with user farms, which are all blade servers with 8 CPU cores per server, form a computing cluster. The CPUs of pqm1, farm1, farm2 are Intel (R) Xeon (R) E5506 @ 2.13 GHz, and those of pqm2 and other user farms are Intel (R) Xeon (R) E5420 @ 2.50 GHz. The Portable Batch System (PBS) is implemented to allocate computational jobs among the available computing resources. The 16 cores of the PQM servers are dedicated to the PQM jobs, which are sufficient during normal physics data taking. And the 40 cores of the user farms can be shared with PQM jobs, in case of some special data taking with higher event rate. The function of other servers will be described in Section~{\ref{sec:dataflow}}.

%\begin{figure}[htb]
\begin{center}
\includegraphics[width=7cm]{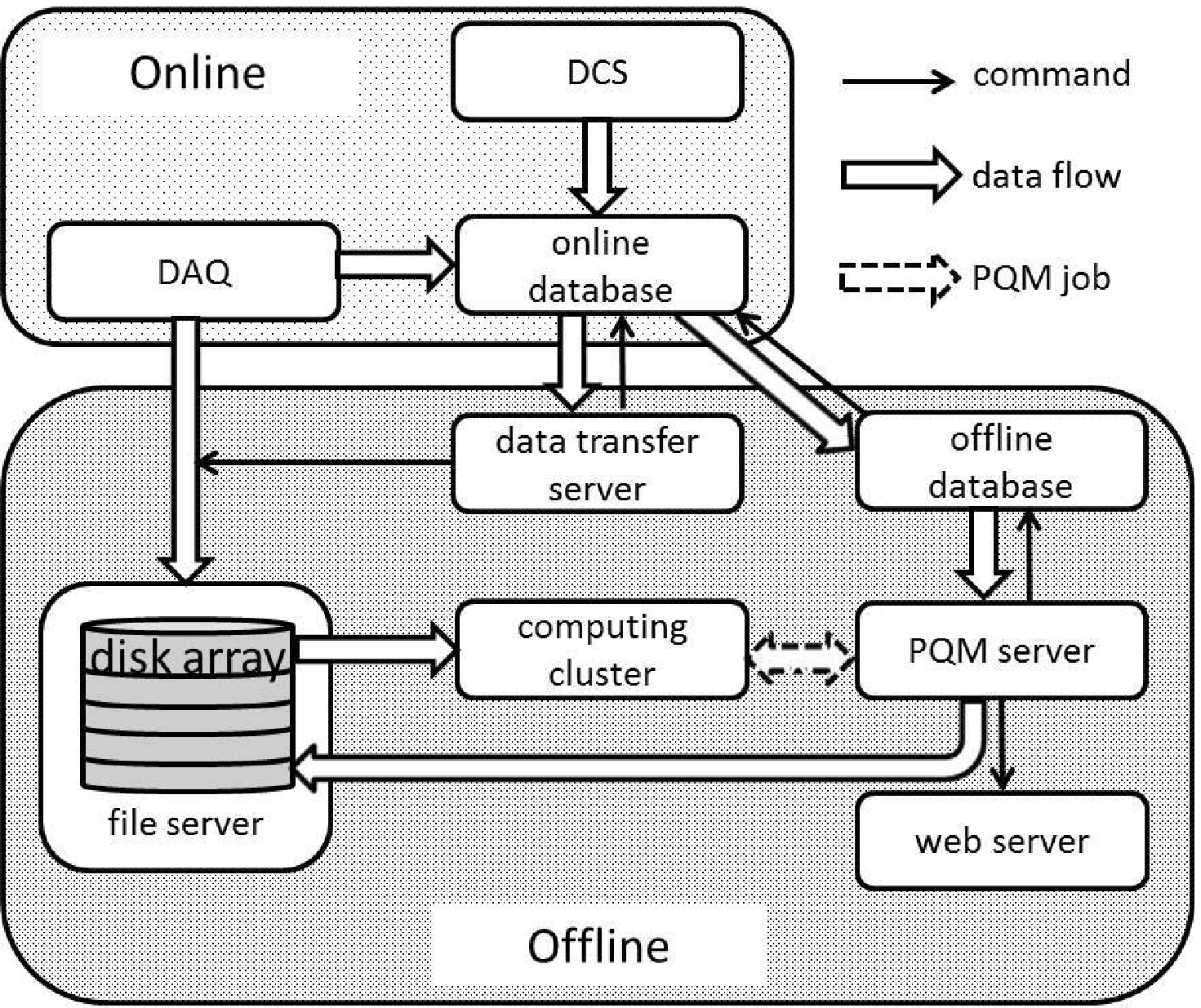}
\figcaption{The schematic diagram of the offline computing environment at Daya Bay.\label{fig:oe}}
\end{center}
%\end{figure}

The total disk volume of the file server is around 25 Terabytes (TB). 17 TB are for storing raw data files from the DAQ, 5 TB for onsite users, 2 TB for offline softwares and 1 TB for storing ROOT~\cite{root} files and histogram pictures produced by PQM jobs.

\subsection{Data flow}\label{sec:dataflow}

The offline computing environment communicates with the online systems to process onsite data. Here we briefly introduce the related online system parts, as shown in Fig.~\ref{fig:oe}, which are the DAQ system, the Daya Bay Detector Control System (DCS), and the online database. When the DAQ closes a raw data file, the information, including the starting and end timestamps of the file, the run number, the file number and so on, are dumped into the online database to generate a new record. The values of high voltages of the photo-multiplier tubes (PMTs) and the RPC modules, the detector temperature, the water levels of the water shields, etc., which are provided by the DCS, are also recorded in the online database.

To prevent interruptions to the DAQ, a raw data file is copied to the offline file server after the data transfer server detects a new record in the online database.  Copies of raw data files are stored in the file server for more than 1 month. The offline database server is responsible for copying the information that is necessary for offline analysis from the online database to the offline database. Besides the online information, there are calibration constants in the offline database, such as PMT gains, PMT time offsets, and energy scales of the ADs, which are updated after careful validation. The PQM server is responsible for the implementation of the functionality of the PQM for monitoring the detector performance and the DQ. This will be described in detail in Section~{\ref{sec:implementation}}. The web server is used to display user-defined histograms in analysis algorithm modules.

\subsection{Offline software}

The Daya Bay offline software (NuWa)~\cite{ad12} \mbox{using} Gaudi~\cite{gaudi} as the underlying software framework is available in the offline computing environment. NuWa, which employs Gaudi's event data service as the data manager, provides the full functionality required by simulation, reconstruction and physics analysis. Job modules are used to configure simulation and analysis tasks. Specifically, job modules are scripts which can add analysis algorithms and tools to the job, and configure algorithms, tools, and services used by the job. In the NuWa framework, reconstruction algorithms have been developed to reconstruct the energy and the vertex of the antineutrino event from the charge pattern of the PMTs. One can find the details of the calibration and reconstruction methods in~\cite{ad12}. By default, the latest calibration constants in the offline database are loaded into the reconstruction algorithms, reflecting the best understanding of the detectors.

The auto-building system of NuWa is implemented using the Bitten~\cite{bitten} plug-in of trac~\cite{trac}, which is an enhanced wiki and issue tracking system for software development projects. One of the user farms serves as the Bitten slave and builds NuWa automatically when the code is updated. A series of pre-defined testing jobs follow the building for the purpose of validation. The validated build can then be manually set for PQM jobs if it contains a necessary upgrade of reconstruction or analysis algorithms.

\section{Implementation of the PQM}\label{sec:implementation}

The PQM is designed to efficiently process all events in all raw data files as soon as possible. It communicates with the offline database to get the latest raw data file list and submits jobs to the PBS to process them in parallel. At the end of each job, some selected histograms are printed for web display. The shift crew can compare the new histograms with standard ones to report sub-detector status and the DQ.

\subsection{Data processing chain}

To manage the data flow of the PQM as shown in Fig.~\ref{fig:dataflow}, a control script in Python language has been developed and runs in a background mode on the PQM server. When the DAQ closes a raw data file, the offline database will be updated with a new record copied from the online database. A copy of the file will be transferred to the file server in less than 5 minutes, assuming a size of ($\sim$1 GB). When the data transfer is completed, the status of this file is set as TRANSFERRED in both online and offline database.
\end{multicols}

\ruleup
\vspace{1.0cm}
\begin{center}
\includegraphics[width=14cm]{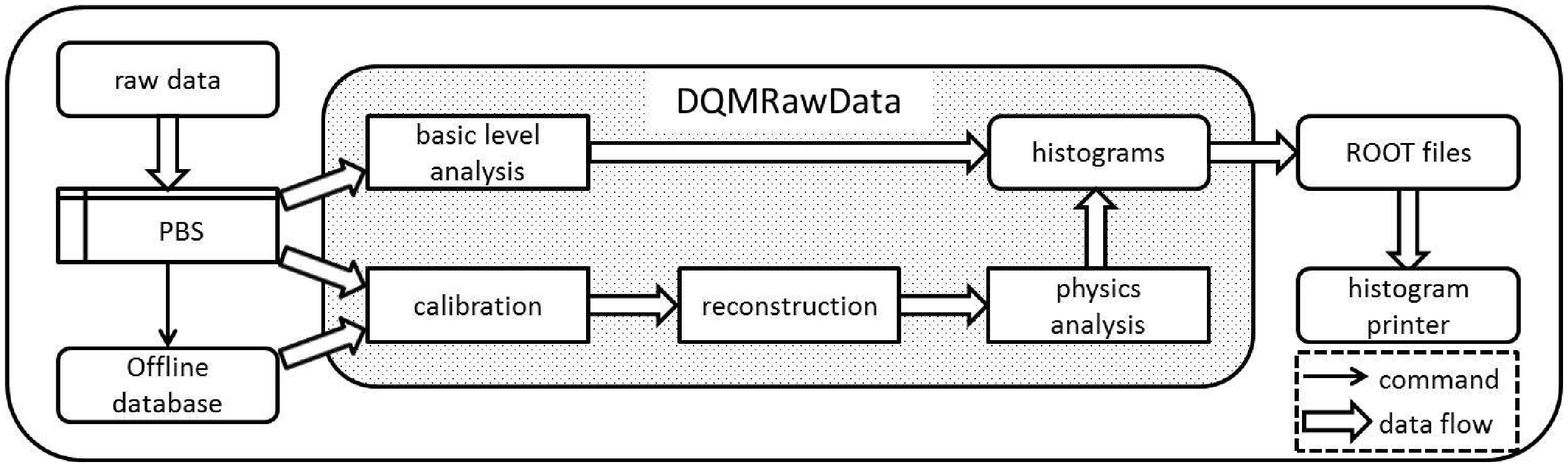}
\figcaption{The data flow of the PQM job.\label{fig:dataflow}}
\end{center}
\vspace{1.0cm}
\ruledown

\begin{multicols}{2}
The control script queries the offline database at a fixed time interval (10 seconds). If a new record is found and the data status returned by the query shows that the corresponding raw data file is already on the file server, a job for processing the new file is submitted to the PBS. The PQM job uses a recently manually tagged NuWa version to reconstruct data with the latest calibration constants in the offline database, runs analysis algorithms, and fills the results into user-defined histograms. All of the histograms are dumped into a ROOT file at the finalization step of the algorithms, which is then merged with the accumulated ROOT file for the same run. After that, a histogram printer in C++ language is executed taking the merged ROOT file as input to print selected histograms into figures. The control script also checks if the job is finished at a 10-second interval by detecting an empty TXT file created at the end of the job. For finished jobs, the control script transfers the corresponding figures to the PQM disk for web display (see Section~{\ref{sec:webdisplay}}). When the run ends and all the corresponding raw data files are processed, the control script saves the accumulated ROOT file on the PQM disk for permanent storage.

\subsection{Analysis modules in PQM}

Currently four modules are developed for the PQM to analyze data and fill user-defined histograms in the NuWa framework: one for histograms of electronics channel information, one for the histograms of calibrated PMT information of ADs and the water shields, one for the histograms of the reconstruction for the ADs, and the other one for the histograms of the RPCs. The histograms displayed by the web interface are shown in Table \ref{tab:histlist}. The detailed description of the variables listed in Table \ref{tab:histlist} can be found in~\cite{ad12,rpc_off,rpc_calib}. It provides the flexibility to add more algorithm modules as well as to fill more user-defined histograms.

Since a different configuration of detectors is adopted for EH3, all the analysis modules take the strategy of creating histograms dynamically for each detector, and even for each PMT channel to reduce the output file volume and to improve the processing efficiency. If one event read from the raw data file is the first one for the current detector, a new set of histograms for this detector are created. Otherwise, the event is analyzed and filled into existed histograms.
\end{multicols}

\ruleup
\vspace{1.0cm}
\begin{table}[htb]
\begin{center}
\footnotesize
\begin{tabular}{llp{11.5cm}}
   \toprule
  Detector Unit        & Level & Histograms \\\hline
  PMT                  & Basic & Mean of ADC, TDC and preADC, RMS of ADC, TDC and preADC, $\Delta$ADC, dark noise, dark rate and TDC vs channel ID, hit rates, ADC sum vs. trigger types, ADC sum, number of blocked triggers vs. run time, number of channels, trigger types\\\cline{2-3}
                       & Reconstruction & 2-dimension vertex distribution, distribution of vertex vs. energy, distribution of event rate vs. vertex, distribution of energy\\\hline
%                       & Reconstruction & y vs. x, z vs. radius, energy, energy vs. radius, energy vs. z, event rate vs. radius, event rate vs. z\\\hline
  RPC                  & Basic & Map of patch rate, map of module rate, map of channel multiplicity, distributions of time between module and system triggers, maps of layer efficiency, distribution of layer efficiency, electronics error types vs. time, maps of channel hit count, channel hit count vs. channel ID, distribution of number of layers per readout, distribution of number of modules per system trigger, maps of layer singles rate, distribution of layer singles rate, module and system trigger rates vs. time.\\
   \bottomrule
  \end{tabular}
  \setlength{\abovecaptionskip}{10pt}
  \caption{The histograms produced by analysis modules in PQM for monitoring the detector performance and the DQ. \label{tab:histlist}}
\end{center}
\end{table}
\vspace{1.0cm}
\ruledown

\begin{multicols}{2}

\section{Performance and information display}\label{sec:webdisplay}

The PQM is required to provide more information about the running status of all sub-detectors in 3 EHs than the online histogram presenter provided by the DAQ. The data transfer from the DAQ takes less than 5 minutes. The job can be submitted to the PBS about 10 seconds after the data transfer is finished. Running NuWa job modules takes around 30 minutes. In the meanwhile, some user-defined histograms produced by job modules are saved in ROOT files. They are printed and published via the web interface in around 2 minutes. The job failure rate is less than 1\% during nomarl data taking. The typical latency of the display of the PQM plots is 40 minutes after a new raw data file is closed by the DAQ. The most time-consuming part is running the calibration and reconstruction algorithms. Since all algorithms are configurable in the PQM job, it is easy to remove them and only show the basic-level information of the detectors to reduce the latency to less than 20 minutes. This special configuration of PQM jobs was usually implemented during the commissioning period, since the onsite experts could see the detector status as soon as possible.

\begin{center}
\includegraphics[width=3.5cm]{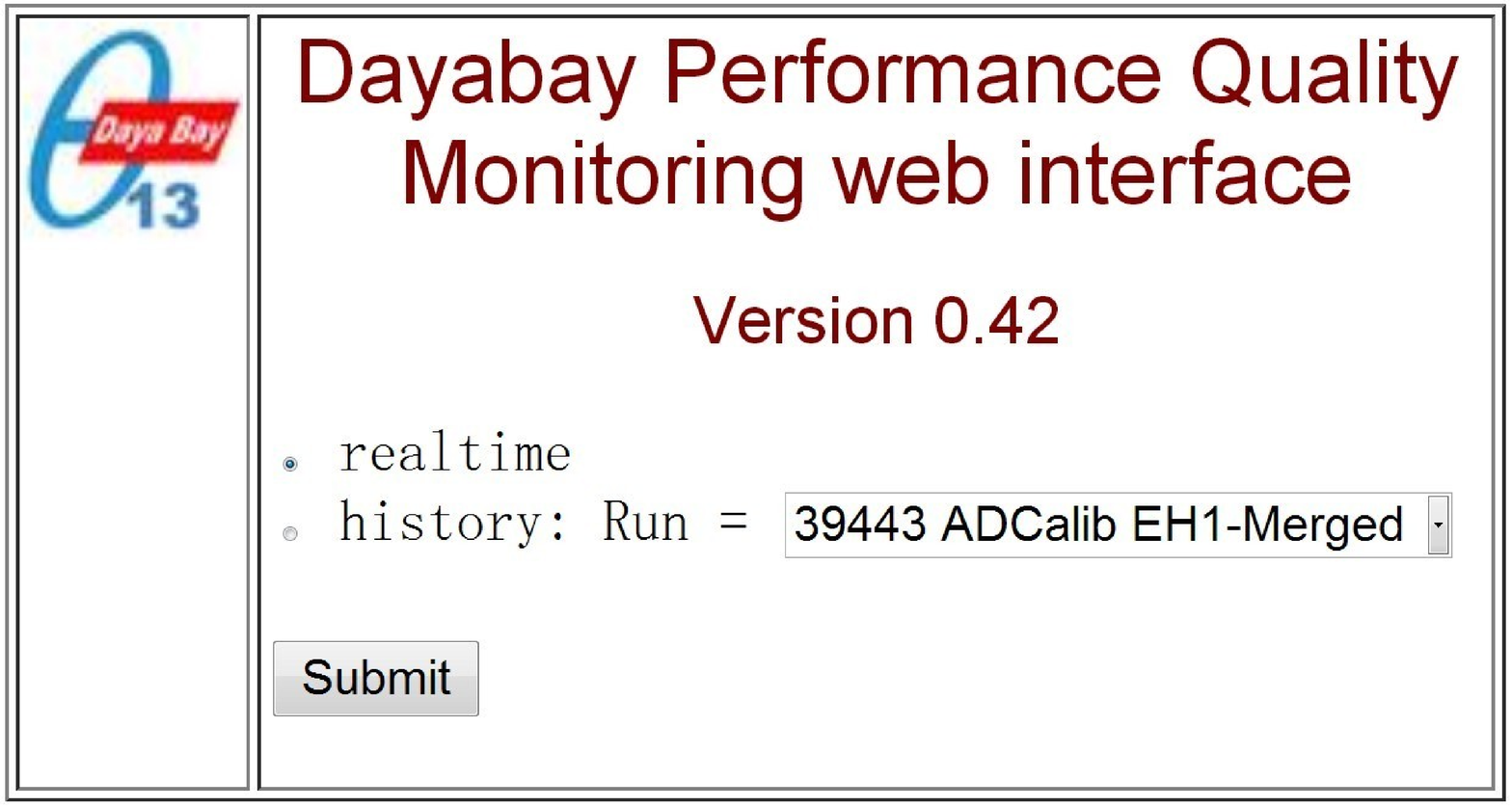}
\hspace{3ex}
\includegraphics[width=2.5cm]{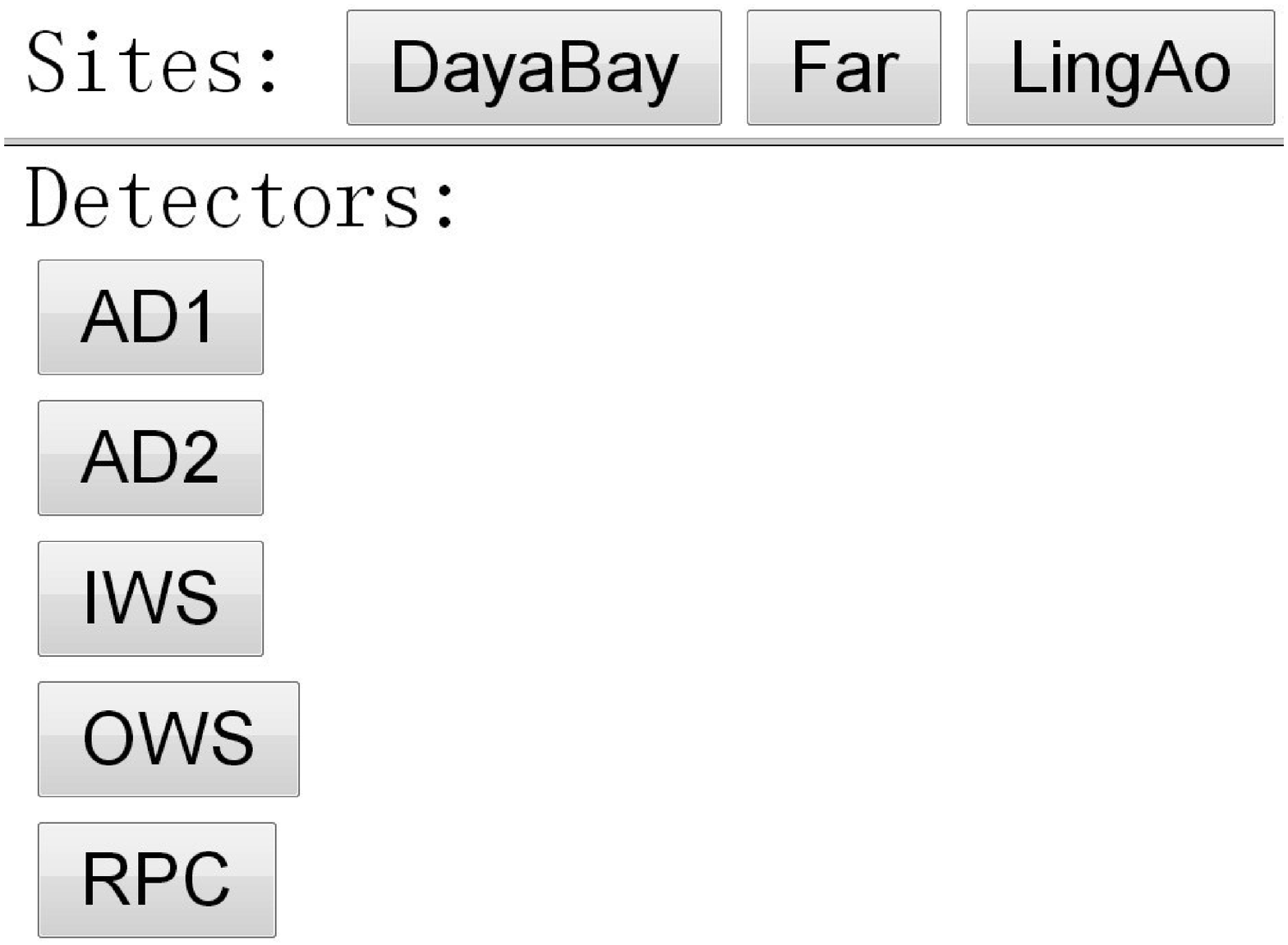}
\figcaption{Web interface for the PQM.\label{fig:web}}
\end{center}

A web interface, as shown in Fig.~\ref{fig:web}, running on the web server is utilized to display the selected user-defined histograms. On the main page (the left panel of Fig.~\ref{fig:web}), two options are provided for users: one is for showing histograms for the current runs of each site, named as `realtime', the other one is for all the previously processed runs, named as `history'. For the latter, one can choose the run number in a drop-down menu listing the site ID and run type, i.e. `Physics', `FEEDiag' and so on. On the `realtime' page (the right panel of Fig.~\ref{fig:web}), site buttons are available for different sites: `DayaBay', `LingAo' and `Far'. The `realtime' page refreshes at a fixed time interval of 1 minute. For each site, the user can also use detector buttons to view histograms for different sub-detectors. On the `history' page, there are only detector buttons, because this page is for the run selected by the user, and the specific run is for only one site.

%\begin{figure}[htb]
\begin{center}
\includegraphics[width=7cm]{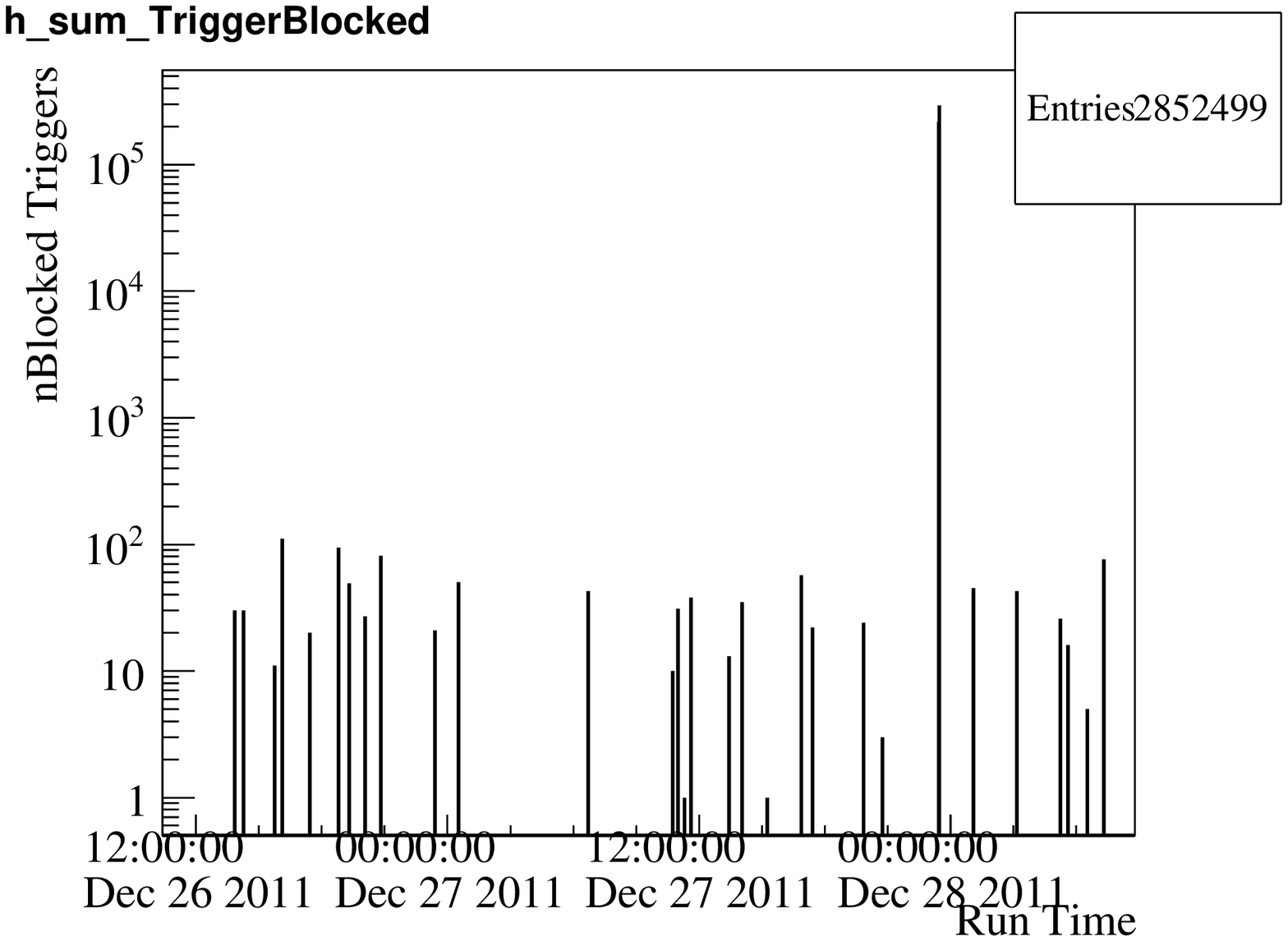}
\figcaption{The number of blocked triggers in one event vs. run time.\label{fig:nblocked}}
\end{center}
%\end{figure}

Several example histograms are given here. Fig.~\ref{fig:nblocked} shows the number of blocked triggers in one event as a function of time. Generally, if a cosmic-ray muon passes through the detector and induces a shower with extremely large energy deposit, overflow of the electronics occurs, and the local trigger and the corresponding data package can be blocked. The blocked trigger number is recorded and read out. If the fraction of the blocked trigger number over the total trigger number is larger than a pre-defined value, the shift crew should report to the DQ working group (DQWG).

Fig.~\ref{fig:energy} presents the reconstructed energy spectrum for all triggers for an AD. The peaks of $^{40}$K (1.46 MeV), $^{208}$Tl (2.61 MeV), and the gamma peak from neutron capture on Gd ($\sim$8 MeV) are labeled in fixed positions in the histogram. The red line for $^{208}$Tl is set to $\sim$2.75 MeV due to the energy non-linearity which is not corrected in the analysis algorithm. A relative shift between the red lines and the observed peaks indicates problems in the data, e.g. incorrect energy scales read from the offline central database or unexpected behaviour of the PMTs.

%\begin{figure}[htb]
\begin{center}
\includegraphics[width=7cm]{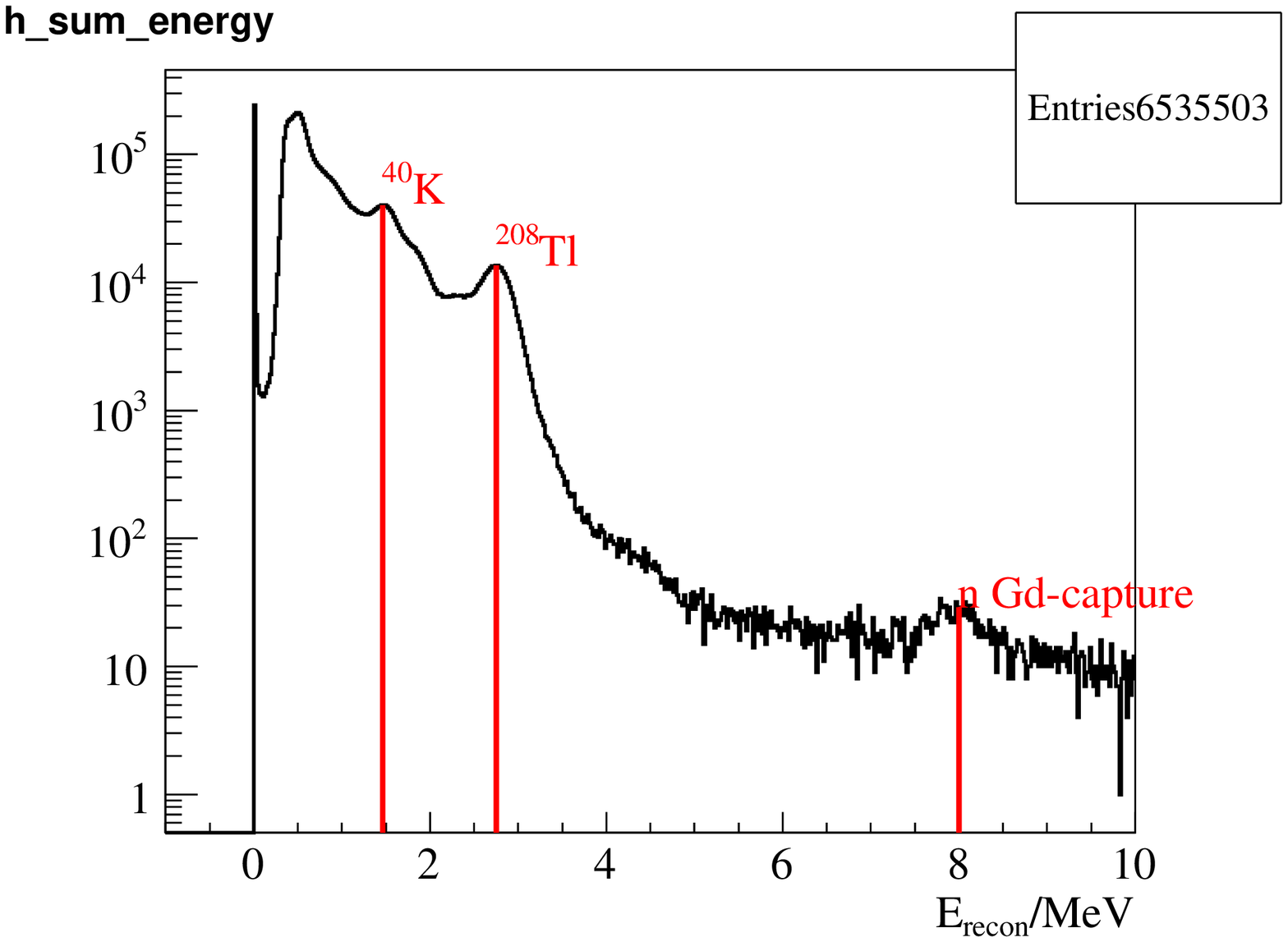}
\figcaption{The reconstructed energy distribution of all events for AD1.\label{fig:energy}}
\end{center}
%\end{figure}

%\begin{figure}[htb]
\begin{center}
\includegraphics[width=7cm]{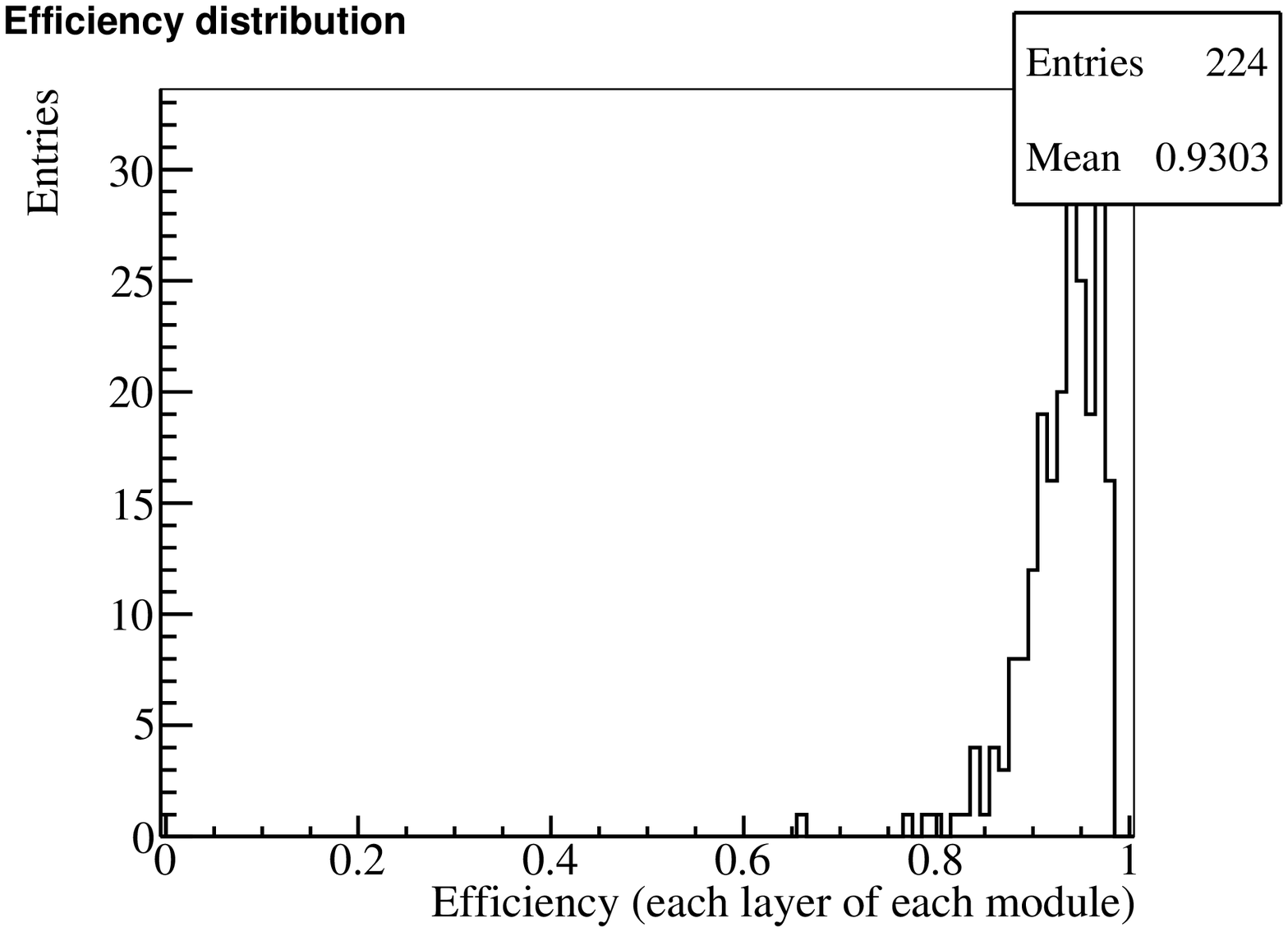}
\figcaption{The efficiency of each layer for the RPCs.\label{fig:rpceff}}
\end{center}
%\end{figure}

The statistics of the efficiency of each RPC layer~\cite{rpc_off,rpc_calib} is displayed in Fig.~\ref{fig:rpceff}. The efficiency is calculated in the algorithm of analysis modules for the PQM. The efficiency is a critical quantity to evaluate the performance of the RPCs used as an anti-coincidence detector for muons for the experiments. As shown in the figure, most of RPC layers are working with the efficiency of about 90\%. The performance of the RPCs can change due to the gas pressure or other unexpected reasons. The RPC experts and the DQWG are informed if the RPC layer efficiency distribution distorts greatly by comparing with the standard ones.

There are also many other histograms to monitor the detector performance and the DQ. The shift crew can compare them with the corresponding standard figures to check the running status of the sub-detectors and the DQ. In the case that irregularities are observed, the shifter should inform the related experts and report to the DQWG.

A supernova monitoring program is under development to join the Supernova Early Warning System (SNEWS)~\cite{snews}, in which individual neutrino-sensitive experiments send supernova (SN) burst datagram to SNEWS coincidence computer at Brookhaven National Laboratory (BNL). Supernova burst neutrinos are important in studying the dynamics of core collapse and supernova explosion, neutrino properties, and many other interesting studies. The three-hall configuration design at Daya Bay can significantly reduce false SN alarms caused by the spallation background. We will have online and offline triggers for monitoring SN bursts. Given the short latency time of the PQM, the offline trigger for SN bursts will be added to the PQM job.

\section{Conclusions}

\par
The onsite data processing is running smoothly. The PQM, which has been developed for the Daya Bay experiment, has the ability to provide user-defined histograms to the shift crew with a latency time of about 40 minutes. It satisfies the requirement for the data taking. By comparing the histograms with the standard ones, potential DQ problem can be found. The PQM will contribute to the SN monitoring program in the Daya Bay experiment in the near future.

\acknowledgments{This work is supported by the National Natural Science Foundation of China (Y2118M005C). The author, LIU Ying-Biao, would like to thank LI Gao-Song for the contribution to the ReconDataHistogram, Logan Lebanowski and ZHANG Qing-Min for the contribution to the RpcDataHistogram, and JI Xiang-Pan for the contribution to CalibDataHistogram.}

\end{multicols}

\vspace{-1mm}
\centerline{\rule{80mm}{0.1pt}}
\vspace{2mm}
\begin{multicols}{2}

\end{multicols}

%\end{CJK*}
\end{document}